\documentclass[twocolumn,prl,showpacs]{revtex4}
\usepackage{graphicx}

\begin{document}

\newcommand{\beq}{\begin{equation}}
\newcommand{\eeq}{  \end{equation}}
\newcommand{\bea}{\begin{eqnarray}}
\newcommand{\eea}{  \end{eqnarray}}
\newcommand{\bit}{\begin{itemize}}
\newcommand{\eit}{  \end{itemize}}

\title{Decoherence and the quantum-classical limit in the presence
of chaos.}

\author{F. Toscano,  
R. L. de Matos Filho, and L. Davidovich.}

\affiliation{Instituto de F\'{\i}sica,
         Universidade Federal do Rio de Janeiro,  Caixa Postal 68.528, 
	 21.941-972, Rio de Janeiro, Brazil}

\date{\today}

\begin{abstract}

We investigate how decoherence affects the short-time separation between quantum and classical 
dynamics for classically chaotic systems, within the framework of a specific model. For a wide 
range of parameters, the distance between the corresponding phase-space distributions depends 
on a single parameter $\chi$ that relates an effective Planck constant $\hbar_{\rm eff}$, the 
Lyapunov coeffficient, and the diffusion constant. This distance peaks at a time that depends 
logarithmically on $\hbar_{\rm eff}$, in agreement with previous estimations of the separation 
time for Hamiltonian systems. However, for $\chi\lesssim 1$, the separation remains small, going 
down with $\hbar_{\rm eff}^2$, so the concept of separation time loses its meaning.   
\end{abstract}

\maketitle

\indent
One of the most subtle problems of quantum mechanics is the 
description of the classical world, particularly for classically 
chaotic systems. Even for initial states that are classically allowed, one expects 
that the dynamics of the quantum and the corresponding classical 
system should differ, after some time. Indeed, while for linear 
systems the Wigner distribution can be shown to obey the same 
dynamical equation than the classical phase-space distribution, non-linearities 
will eventually set the two distributions apart. 

For a classically chaotic system, the separation time can be very 
short, due to the exponential stretching of the distribution, for 
positive Lyapunov coefficients, which quickly  allows the 
distribution to explore the non-linearities of the system, even if the 
linear dimensions of the initial wavepacket are smaller than the typical nonlinear scale of the problem. 
The separation time in this case has been shown by many authors
\cite{separation_time,Zaslasky1991} to scale as $\ln(1/\hbar_{\rm eff})$, 
where $\hbar_{\rm eff}=\hbar/S$, and $S$ is a typical action of the system. 
Thus, even when $\hbar_{\rm eff}\ll 1$, the separation 
time can still be small, as compared to a typical evolution 
time of the system. 
Such logarithmic law may pose a problem to the quantum-classical correspondence 
of macroscopic objects, leading to consequences that contradict observation \cite{Zurek}. Reconciliation of theory and 
observation is provided by the irreversible coupling of the system 
with a reservoir, which leads to the elimination of quantum 
signatures. This has been investigated numerically and analytically for several models \cite{Zurek,Enviroment_influence,Pat}. 

In the presence of the environment, one expects that the logarithmic law should not hold anymore. In Ref.~\cite{Andre}, it was shown that, for the kicked harmonic oscillator, diffusion helps to decrease the difference between quantum and classical variances, so that, if one defines any arbitrary value of this difference as the ``critical separation value'', the separation time becomes infinite for a sufficiently large diffusion coefficient, that is, the difference remains always smaller than the critical value. In that work, however, the dependence on the relevant parameters of the difference between classical and quantum dynamics could not be elucidated.

In this paper, we derive for the kicked harmonic oscillator \cite{Zaslasky1991,Zaslavsky_book} the precise dependence of the separation between quantum and classical distributions on the parameters that  control macroscopicity, noise, and chaotic behavior. These factors, for a wide range of their values, can be combined in a single parameter, which will be shown to govern the quantum-classical transition, as conjectured in Ref.~\cite{Pat} for general chaotic systems. 

The relevant dimensionless parameters correspond to the diffusion coefficient $D$, the kicking strength $K$, and an effective Planck's constant $\hbar_{\rm eff}$. We show that, in the chaotic regime, for any finite diffusion coefficient, and in the semiclassical limit $\hbar_{\rm eff}\ll1$, the distance between the two distributions, defined as the integral of the magnitude of their difference over all phase space, is proportional to $\chi=K\hbar_{\rm eff}^2/4D^{3/2}$, as long as $\chi \lesssim 1$. In this regime, the time for which this distance peaks is shown to be a logarithmic function of $\hbar_{\rm eff}$, as estimated previously for Hamiltonian chaotic systems. However, in this case the concept of separation time is not meaningful anymore, since the two distributions remain close together throughout the evolution. 

The Hamiltonian of the kicked harmonic oscillator is defined as:
\beq
\label{kho}
\hat{H}=\frac{\hat{P}^2}{2m}+\frac{1}{2}m\nu^2\hat{Q}^2+
A\;\cos(k\hat{Q})\sum_{n=0}^\infty \delta(t-n\tau).
\eeq
This Hamiltonian can be shown to describe the center-of-mass dynamics
of an ion in a one-dimensional trap submitted to a sequence of
standing-wave laser-pulses, off-resonance with a transition between
the electronic ground state and another internal state 
\cite{Gardiner1997}. The wave number $k$ in
Eq.~(\ref{kho}) is the projection along the trap 
axis of the corresponding (identical) wave vectors of two opposite
propagating pulses with oblique incidence. The high degree of control in ion experiments \cite{Wineland}, plus the possibility of engineering several kinds of reservoir for the center-of-mass motion \cite{Wineland2}, greatly stimulates the interest in using this system for testing fudamental features of the quantum-classical transition. The corresponding phase space is unbounded, and no long-time localization occurs, as opposed to the kicked rotator, which has been subject to experimental test concerning the dependence of localization on noise \cite{Raizen}.  

It is convenient to work with the dimensionless quantities
$\hat{q}=k\hat{Q}$, $\hat{p}=k\hat{P}/m\nu$, and $K=k^2A/m\nu$,
so that $[\hat{q},\hat{p}]=2i\eta^2\equiv i\hbar_{\rm eff}$, with
$\eta=k\Delta Q_0=k\sqrt{\hbar/2m\nu}$, and $\Delta Q_0$ being the width of the ground state of the harmonic
oscillator. 
The dimensionless parameter $\eta$ is the so-called Lamb-Dicke
parameter \cite{Wineland}, which measures the ratio between the 
ground state width and the wavelenght $\lambda=2\pi/k$ that sets 
the scale of the non-linearity of the Hamiltonian. 
It is important to note that, in experiments with trapped ions, 
the classical limit $\eta \rightarrow 0$ can be approximated 
simply by changing the angle of incidence of the incoming pulses,
or by increasing the trap frequency.  

The time-dependent classical evolution can be described as a composition 
of a discrete map corresponding to the kick, plus a rotation in 
phase space:
${\bf x}_{n+1}={\bf R}\circ {\bf K}({\bf x}_{n})$ where
${\bf x}_n\equiv(q_n,p_n)$ are the coordinates before  
 kick $n$ (note that our first kick corresponds to $n=0$). The operation ${\bf K}$ is defined as
\beq
\label{K_operation}
q_{n}^+=  q_{n}\,,\qquad
p_{n}^+= p_{n} + K\,\sin(q_n)\,,
\eeq
and the phase space rotation ${\bf R}$ is given by:
\bea
\label{R_operation}
 q_{n+1}&=& \cos(\nu\tau)\,q_{n}^+ + \sin(\nu\tau)\,p_n^+ \,,\nonumber\\
p_{n+1}&=& -\sin(\nu\tau)\,q_{n}^+ + \cos(\nu\tau)\,p_n^+\;.
\eea
For definiteness, we consider here  
$\tau=T/6$ ($T=2\pi/\nu$ is the period of the harmonic oscillator)
wich leads to the so-called ``stochastic web'' in phase space
\cite{Zaslasky1991}, {\it i.e.} a pattern of groups of stability 
islands with hexagonal symmetry immerse in a sea of chaotic 
trajectories. The strong-chaos regime corresponds to $K\gg 1$.  

\begin{figure}[b]
\setlength{\unitlength}{1cm}
\begin{picture}(8.0,6.8)(0,0)
%\put(0.95,4.95){\includegraphics[angle=0,width=6.7cm]{fig1a.ps}}
\put(0.95,4.95){\includegraphics[angle=0,width=5.2cm]{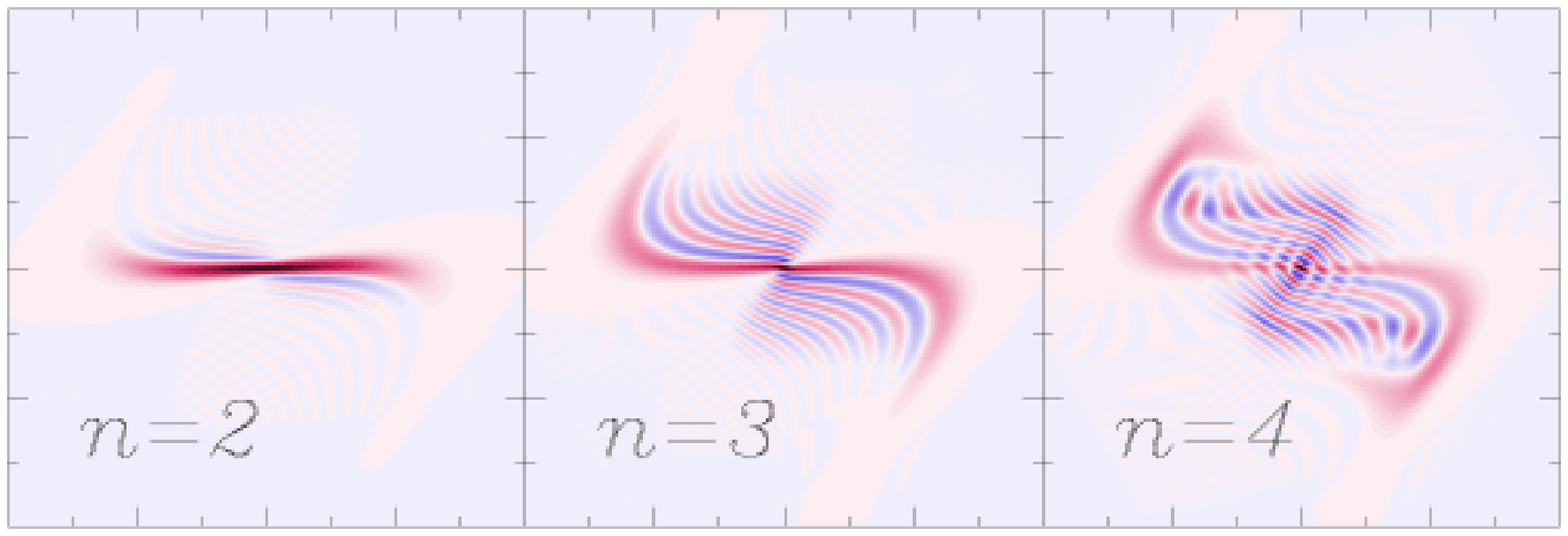}}
\put(2.6,0.6){\includegraphics[angle=0,width=5.2cm]{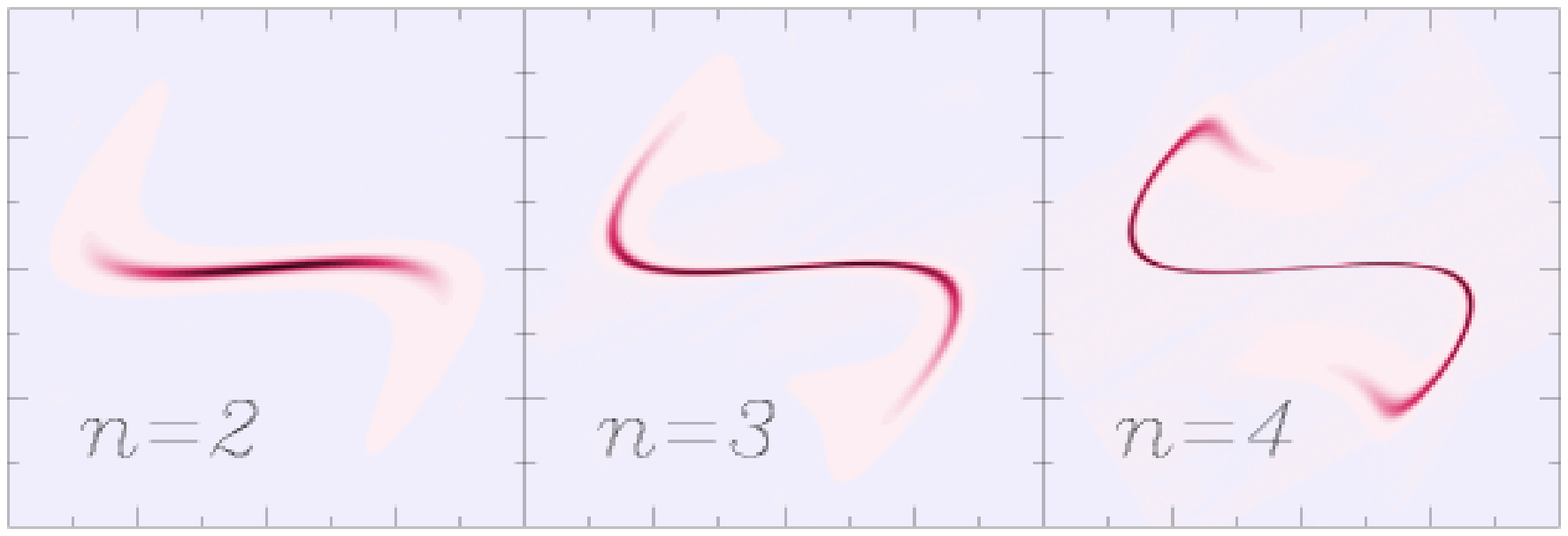}}
\put(0,5){\includegraphics[angle=-90,width=7.2cm,bb=136 98 539 624]{fig1b.ps}}
\end{picture} 
\vspace*{1.0pc}
\caption{(Color online) Separation ${\mathcal D}_n$ between quantum and classical distributions as function of the number of kicks, for $\eta=0.3$, $K=2$, for unitary evolution.  The initial distribution is a coherent state centered around the origin of phase space with width $\Delta q=\Delta p=\eta$ much smaller than the nonlinearity scale ($\eta\ll 1$). The peak correponds to the first folding of the distribution, leading to the appearance of interference fringes in the Wigner function (upper inset). The lower inset displays the classical distribution. The folding appears when the distribution starts probing the nonlinearity.}
\label{fig1}
\end{figure}

In order to study the separation between quantum and classical dynamics, we shall use the quantity  
\beq
\label{functional_distance}
{\mathcal D}_n\equiv 
\int d{\bf x} 
\left|W_{n}({\bf x})-W_{n}^{cl}({\bf x})\right|\;.
\eeq
where $W_n$ and $W_{n}^{cl}$ are respectively the Wigner and the classical distributions immediately before kick $n$. The normalization of the distributions is taken equal to one. A typical example of the variation of ${\mathcal D}_n$ as a function of the number of kicks, in the absence of decoherence, is shown Fig.~\ref{fig1}. Separation becomes evident as ${\mathcal D}_n$ gets larger than one. The first peak of ${\mathcal D}_n$ corresponds to the first folding of the distributions, as their widths become of the order of one and the system starts exploring the nonlinearity of the Hamiltonian.  

%%%%%%%%%%%%%%%%%
%    FIGURE 1
%%%%%%%%%%%%%%%%%

We discuss now how decoherence affects the behavior of ${\mathcal D}_n$. The effect of a thermal reservoir with average population $\bar{n}$, in the Markovian and weak-coupling limit, may be described by the Fokker-Planck equation for the Wigner function
\begin{eqnarray}
\label{Fokker_Planck}
\left. {\frac{{\partial W}}{{\partial t}}} \right|_{{\rm{reservoir}}} &=& \frac{\Gamma }{2}\left[ {\frac{\partial }{{\partial q}}\left( {qW} \right) + \frac{\partial }{{\partial p}}\left( {pW} \right)} \right]\nonumber \\
[0.5ex]&+& \Gamma \left( {\bar n + \frac{1}{2}} \right)\eta ^2 \left( {\frac{{\partial ^2 W}}{{\partial q^2 }} + \frac{{\partial ^2 W}}{{\partial p^2 }}} \right)\,,
\end{eqnarray}
where $\Gamma$ is the dissipation rate. For the complete evolution we must add to the right-hand side 
(r.h.s.) of Eq.~(\ref{Fokker_Planck}) the unitary evolution given 
by Eq.~(\ref{kho}).
In the classical case, the complete evolution is obtained by adding the corresponding Liouville term
to the r.h.s. of Eq.~(\ref{Fokker_Planck}).

In the low-temperature limit $\bar{n}\ll 1$, the diffusion term becomes negligible in the semiclassical regime $\eta\ll 1$, and one gets then purely dissipative dynamics. On the other hand, when $\bar{n}\rightarrow\infty$, $\Gamma\rightarrow0$, with $\bar{n}\Gamma$ constant, one gets a purely diffusive dynamics. In this limit, Eq.~(\ref{Fokker_Planck}) is also the diffusion equation for a classical distribution $W^{cl}({\bf x})$, as long as $\bar{n}\Gamma\eta^2$ is identified with a classical diffusion constant $\tilde\Gamma$.

The first limit has already been investigated in Refs.~\cite{Zaslavsky2003,Andre}, where it was shown that pure dissipation, in the sense described above, does not change the logarithmic law for the separation time. In fact, they show that the separation time ${t}_S$, for an initial wave packet centered around the origin of phase space, is given by
\begin{equation}\label{ts}
t_S\approx\frac{\tau}{\lambda}\ln(1/\eta)\,,
\end{equation}
where $\lambda$ is the logarithm of the expansion eigenvalue of the linearized map at the origin.  With dissipation, the expansion eigenvalue includes a factor $\exp(-\Gamma \tau/2)$, so in this case $\lambda=\lambda_{0}-\Gamma\tau/2$, where $\lambda_0$ is the expansion eigenvalue without dissipation. If instead one takes an average over initial conditions in the chaotic region, then one can show that $\lambda$ is replaced by the Lyapunov coefficient $\Lambda$.  With dissipation, one has $\Lambda=\Lambda_{0}-\Gamma\tau/2$. The limit $\Gamma\tau/2=\Lambda_{0}$ is never
attained in the chaotic regime, since before that the chaotic behavior disappears, 
being replaced by simple attractors in phase space \cite{Andre}. 
This implies that dissipation by itself is not 
an efficient mechanism to increase the quantum-classical 
separation time, as defined by Eq.~(\ref{ts}). In fact, the logarithmic dependence remains, 
in spite of the dissipation. For the system considered here, in the strong chaos and weak dissipation regime, $\Lambda_0=\ln\left[(K/2)\sin(\nu\tau)\right]$ and $\lambda_0=\ln\left[K\sin(\nu\tau)\right]$.

We show now that diffusion can drastically change this scenario.
Without a reservoir the unitary evolution of the Wigner 
function can be written as  \cite{Berry1979}
\beq
\label{Wigner_evolution}
W_{n+1}({\bf x})=
\int d{\bf x'}\;L({\bf x}^R,{\bf x'})\;W_n({\bf x'})
\,\,,
\eeq
where $L({\bf x}^R,{\bf x'})$ is the propagator
\beq
\label{quantum_propagator}
\int_{-\infty}^{\infty} \frac{d\mu}{2\pi \eta^2}\;
e^{\frac{i}{\eta^2}\left[K \sin(q')\sin(\mu)-
\mu(p^R-p')\right]}\,
\delta\left(q^R-q'\right)\,,
\eeq
corresponding to one kick plus a harmonic evolution, with
${\bf x}^R\equiv[q^R({\bf x}),p^R({\bf x})]\equiv
{\bf R}^{-1}({\bf x})$ being the 
phase space coordinates rotated with the inverse transformation
of Eq.~(\ref{R_operation}).
The Liouville evolution of the classical distribution 
$W^{cl}_n({\bf x})$ can also be written in the form of 
Eq.~(\ref{Wigner_evolution}) with the classical propagator
\beq
\label{classical_propagator}
L^{cl}\left({\bf x}^R,{\bf x'}\right)\equiv
\delta\left[p'-p^R+K\sin(q')\right]\,\delta\left(q^R-q'\right)\;.
\eeq
In the classical limit $\eta\rightarrow 0$, stationary-phase techniques  \cite{Berry1979} guarantee that this 
classical propagator is formally recovered from 
the quantum one. 

Diffusion leads to a smoothing of the propagator, for both classical and quantum
cases. The classical smoothed propagator becomes
\beq
\label{smooth_classical_propagator}
\tilde{L}^{cl}({\bf x}^R,{\bf x'})=\frac{e^{-\left(x^2+y^2\right)}}{4\pi D}\;,
\eeq
where $y=\left( {p' - p^R  + K\sin q'} \right)/2\sqrt D$ and $x= \left( {q' - q^R } \right)/2\sqrt D$.
 
Note that when $D\rightarrow 0$ we recover 
Eq.~(\ref{classical_propagator}).
In the quantum case diffusion leads to a factor $\exp\left(-D\mu^2/\eta^4\right)$ 
in the integrand of Eq.~(\ref{quantum_propagator}).
When the width of this Gaussian is small, $\eta^2/\sqrt{D}\ll 1$, the $\mu$'s that effectively contribute to 
the integration are those close to the origin.
This allows us to use $\sin(\mu)\approx\mu-\mu^3/6$ in the phase
of the integrand. 
Moreover, if $\chi=K\eta^4/D^{3/2}\ll 1$, the term with $\mu^3$ in 
the phase is small, so we can use 
$e^{i(\theta+\delta)}\approx e^{i\theta}+
i\delta\,e^{i\theta}$ and then perform the $\mu$-integration. One should note that, when $\eta<\sqrt{K}$, which is always true when $K>1$ and $\eta<1$, then $\chi<1$ implies that $D>\eta^4$. We get then the following approximation to the smoothed  
quantum propagator:
\beq
\label{approx_smooth_quantum_propagator}
\tilde L ({\bf x}^R,{\bf x'}) \approx \tilde L^{cl}({\bf x}^R,{\bf x'})\left[1  + \chi
 \sin (q')f(y) \right]\,,
\eeq
where $f\left( y \right) = {1}/{4}\left( y - {2y^3 }/{3} \right)$.

The correction to the classical propagator depends only on $\chi$. Since $\left| {f\left( y \right)}exp\left(-y^2\right) \right| \le 0.081$, Eq.~(\ref{approx_smooth_quantum_propagator}) is valid under the less restrictive condition $\chi\lesssim1$.   One should note that $\chi$ depends on the Lyapunov coefficient through $K$. For $K\gg 1$, the Lyapunov coefficient $\Lambda_0$ depends logarithmically on $K$, and therefore $\chi$ is proportional to $\exp\left(\Lambda_0\right)$, which differs from the general expression conjectured in Ref.~\cite{Pat}.
%%%%%%%%%%%%%%%%%
%    FIGURE 2
%%%%%%%%%%%%%%%%%
\begin{figure}[tb]
\includegraphics[angle=-90,width=8cm]{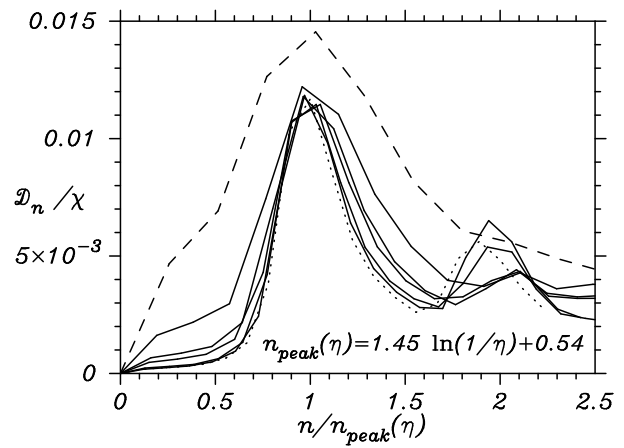}  
\caption{Renormalized distances between quantum and classical distributions, with diffusion. The seven different curves correspond sequentially to  
$\eta=$0.1 (dashed line), 0.04, 0.02, 0.015, 0.007, 0.005, 0.003 
(dotted line) and $D=5.13\times10^{-2}$ (dashed line),
$4.5\times10^{-3}$, $7\times10^{-4}$, $3.25\times10^{-4}$,
$4.5\times10^{-5}$, $1.74\times10^{-5}$, $4.5\times10^{-6}$
(dotted line). In all the cases, $\chi=0.017$. $n_{\rm peak}(\eta)\tau$ is the time at which the first peak of the distance corresponding to a given value of $\eta$ occurs.The second peak happens at the second folding of the distribution.}
\label{fig2}
\end{figure}

The fact that $\chi$ rules the corrections to the classical propagator implies that the separation between the classical and quantum distributions, defined by Eq.~(\ref{functional_distance}), is also scaled by this parameter. This is shown in Fig.~\ref{fig2}, which displays the scaled separation, for a wide range of values of $\eta$ and $D$. The horizontal axis is scaled by $n_{\rm peak}(\eta)$, the number of kicks for which the first peak of ${\mathcal D}_n(\eta)$ is attained.  All the curves fit in the same scale, which establishes that, for $D\ge (K\eta^4)^{2/3}>\eta^4$, which is always attainable in the semiclassical limit, independently of the value of $D$, the separation between the quantum and the classical distributions goes down with $\eta^4$, and therefore may become arbitrarily small. Fig.~\ref{fig2} also shows that the position of the peaks as a function of $\ln{1/\eta}$ is fitted by the straight line $n_{\rm peak}=1.45 \ln \left( 1/\eta\right) + 0.54$, in excellent agreement with Eq.~(\ref{ts}), since in this case $1/\lambda_0=1.47$. Therefore, the time when the first peak in ${\mathcal D}_n$ occurs still behaves logarithmically with $\eta$. However, in view of the smallness of ${\mathcal D}_n$, the concept of separation time is not meaningful anymore. 

\begin{figure}
\centering
\includegraphics[angle=0,width=7.5cm]{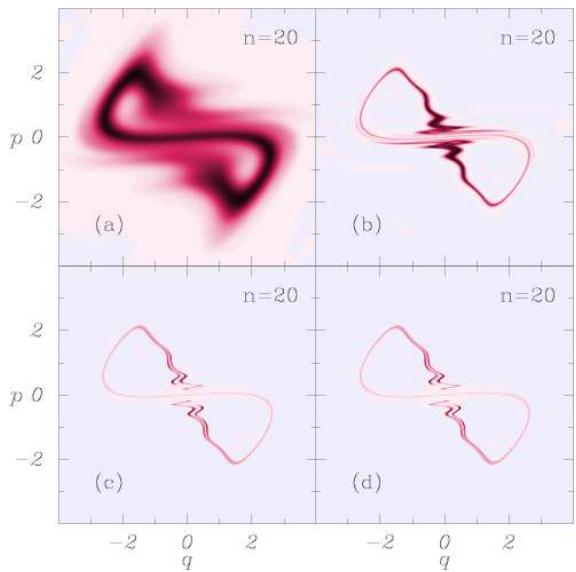} 
\caption{(Color online) Distributions immediately before
the kick $n=20$, for $\chi=0.017$. Classical: {\bf (a)} $\eta=0.04$, $D=4.5\times10^{-3}$, 
{\bf (b)} $\eta=0.007$,  $D=4.5\times10^{-5}$ and {\bf (c)} 
$\eta=0.003$, $D=4.5\times10^{-6}$. Wigner: 
{\bf (d)} same parameters as in {\bf (c)}.}
\label{fig3}
\end{figure}

Note that the condition $D\ge \eta^4$ prevents us from taking the limit $D\rightarrow 0$ for every fixed
value of $\eta$. This restriction does not necessarily imply a diffusion strong enough to substantially change the chaotic classical dynamics,  since we are interested in the semiclassical limit $\eta^2\ll 1$. Figure \ref{fig3} displays the classical and quantum phase-space portraits for $\chi=0.017$ and $K=2$, for different values of $D$ and $\eta$. Since $\chi$ is fixed, as $\eta$ decreases, so does $D$, implying that it is possible, even for $D\ge (K\eta^4)^{2/3}$, to have a non-trivial dynamics, which is not dominated by diffusion. 
%%%%%%%%%%%%%%%%%
%    FIGURE 3
%%%%%%%%%%%%%%%%%

\begin{figure}[t]
\centering
\includegraphics[angle=-90,width=7.5cm]{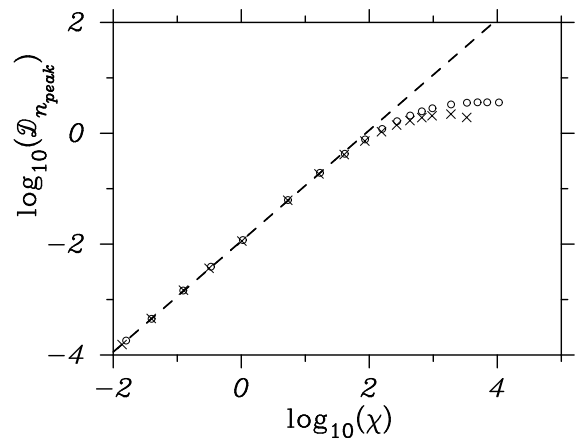} 
\caption{Maximum distance between quantum and classical distributions as a function of $\chi$ for $K=2$, and  $D=4.5\times 10^{-5}$ (circles) or $D=4.5\times 10^{-4}$ (crosses). The linear behavior (dashed line), characterized by the unit value of the slope of the linear region, is verified up to values of $\chi$ well beyond one, for which the distributions are clearly separated.}
\label{fig4}
\end{figure}

Up to now, only the case $\chi\lesssim1$ has been considered. The behavior of the distance for other values of $\chi$  
is displayed in Fig.~\ref{fig4}, which shows that the proportionality to $\chi$ is valid up to ${\mathcal D}_n\approx1$, when the concept of separation time becomes appropriate. 

In conclusion, we have shown that, in the macroscopic limit, and in the chaotic regime, the distance between quantum and classical distributions scales as $\chi=K\eta^4/D^{3/2}$, when $\chi\lesssim 1$, which implies that, no matter how small $D$ is, this distance can be made as small as one wants, by decreasing the effective Planck constant $\eta$.  

We have also shown that the maximum distance between the two distributions is attained at a time that scales logarithmically with $\ln(1/\eta)$. 
Without diffusion, this leads to the well-known logarithmic dependence of the separation time. With diffusion, and $\chi\lesssim 1$, although this behavior of the peaks is still present, its use to define a separation time becomes meaningless. 

This work was partially supported by the Brazilian agencies CNPq, FAPERJ, and FUJB, and the programs PRONEX and Millennium Institute on Quantum Information. We thank A. R. R. Carvalho, C. H. Lewenkopf, A. M. Ozorio de Almeida, and R. Vallejos for discussions.
%%%%%%%%%%%%%%%%%
%    FIGURE 4
%%%%%%%%%%%%%%%%%

\end{document}